# Bistable random momentum transfer in a linear on-chip resonator

Tingyi Gu[1,2†*], Lorry Chang[1†], Jiagui Wu[2,3], Lijun Wu[4], Hwaseob Lee[1], Young-Kai Chen[5], Masudur Rahim1, Po Dong[5], and Chee Wei Wong[2,3]

[1] Department of Electrical Engineering, University of Delaware, Newark, DE 19711, United States

[2] Columbia University, New York, NY 10027, United States

[3] Fang Lu Mesoscopic Optics and Quantum Electronics Laboratory, University of California, Los Angeles, CA 90095, United States

[4] Condensed Matter Physics and Materials Science Division, Brookhaven National Laboratory, Upton, New York 11973, United States

[5] Coherent, 48800 Milmont Drive, Milmont, California 94538, United State

[†]Authors contributed equally

*Corresponding author: Tingyi Gu, Chee Wei Wong

Email:  tingyigu@udel.edu, cheewei.wong@ucla.edu

**Abstract**

Optical switches and bifurcation rely on the nonlinear response of materials. Here, we demonstrate linear temporal bifurcation responses in a passive multimode micro-resonator, with strongly coupled chaotic and whispering gallery modes (WGMs). In microdisks, the chaotic modes exhibit broadband transfer within the deformed cavities, but their transient response is less explored and yields a random output of the analog signal distributed uniformly from '0' to '1'. Here, we build chaotic states by perturbing the multi-mode microring resonators with densely packed silicon nanocrystals on the waveguide surface. In vivo measurements reveal random and 'digitized' output that ONLY populates around '0' and '1' intensity levels. The bus waveguide mode couples firstly to chaotic modes, then either dissipates or tunnels into stable WGMs. This binary pathway generates high-contrast, digitized outputs. The fully passive device enables real-time conversion of periodic clock signals into binary outputs with contrasts exceeding 12.3 dB, data rates of up to $10^7$ bits per second, and 20dB dynamic range.

**Introduction**

Hardware random sequence generators support secure communications [1], quantum simulation [2], probabilistic machine learning [3-4], and regenerative artificial intelligence [5]. The physical entropy sources, such as quantum noise [6-7], electronic shot noise, and chaotic laser noise [8-14], generate random, continuously varying analog signals. These sources typically require precise threshold tuning to minimize the statistical bias, and advanced analog-to-digital conversions (ADC) to interface with digital communication and computing systems (Figure 1a). Single photon sources can directly generate binary quantum random bins in real time [15-17]. Geometric deflection and scattering in resonators disrupt light propagation and circulation, leading to optical loss, mode mixing, and chaotic behavior [18-20]. Chaotic modes in deformed microdisks have been associated with broadband waveguide coupling [21], coherent perfect absorption [22], and directional emission [23-24]. These modes can be engineered to achieve specific spectral and temporal responses [25-26]. However, most studies to date have focused on steady-state excitations [23,27]. MRRs, closely related to microdisks, can be lithographically defined and tailored with high precision to achieve any desired geometry. Geometric engineering enables dispersion control, precision mode selections, and multi-mode operation in MRR [28], but the implementation of chaotic modes has not been reported.

Here we present an alternative approach using a chip-scale integrated passive micro-scale component for directly converting an input periodic clock signal into an optical binary pattern *via* a linear chip-scale integrated MRR (Figure 1b). By designing and fabricating a passive multimode silicon nitride (SiN) microring resonator (MRR) with high-refractive-index silicon nanocrystal (Si NC)-embedded oxide claddings (Figure 1c-g), we achieved a random bit output exceeding $10^7$ bits in a fully passive system. Multimode MRRs were fabricated on a silicon-on-insulator substrate, supporting chaotic modes while retaining the high-*Q* whispering gallery mode (WGM). The waveguide core is made of SiN with a high-refractive index shell of densely packed and randomly distributed Si NCs, inducing mode conversions and chaos within the cavity. We observed a random temporal bifurcation during cavity build-up with pulsed inputs, similar to the recurrent engagement of waves with photonic structures [29]. Unlike conventional chaotic optical signal generators, the output of this analog system is directly binary and insensitive to threshold power (Figure 4a). This binary pattern generation occurs in real-time.

**Results**

Figure 1c is the micrograph of the MRR manufactured through 300mm wafer run [30-31] (Method). A thick (up to 2μm) hydrogenated amorphous silicon oxide (a-SiO$_x$:H) was then deposited on top of the SiN waveguides and MRRs. After high-temperature rapid annealing, Si NCs are found to form on the SiN-oxide interface [32-33]. The measured loaded *Q* is typically ~ 75,000 (linewidth of 20pm) and fiber-chip inverse couplers are included for less than 3-dB coupling loss per facet and total fiber-chip-fiber loss of 14 dB typically (Supplementary Section S1). We used combined micro-photoluminescence (micro-PL) mapping approaches to identify the silicon NCs' distribution (Method). Those silicon NCs were embedded in the oxide matrix surrounding the SiN waveguide, which is identified by the micro-PL mapping (Figure 1d). Tunneling electron microscope (TEM) images identify the average nanocrystal size and density are 3.5 nm and 5×10$^{12}$ cm$^{-2}$, respectively (Method). The effective distance between silicon NCs is estimated to be 4.4 nm. The crystalline cores are consistent with diamond cubic silicon. Silicon (220) plane fringes are in the middle with the adjacent spacing ~ 1.92 Å (Figure 1g). After selectively removing most of the top oxide through wet etching (Figure 1f), the material non-uniformity is verified (which leads to guided modes conversions and chaos). Note the densely packed silicon NCs significantly increase the effective index of the local oxide to be 2.2 (~200nm thick). The hybrid waveguide has ~40 modes and can support chaotic mode with silicon NCs-assisted mode diffusion (representative cross-sectional mode profiles shown in Figure 2b insets) [34] (also in Figure S2-3). The pulsed inputs from the bus waveguide first couple to the high-order chaotic modes (left panel of Fig. 1e) and then either dissipated or converted to the WGM (right panel of Figure 1e). If the involved mode is far away critical coupling point, minimal attenuation of the transmitted light pulse is expected (represented as "1" state on the transmitted through port). The other mode is close to the critical coupling point, resulting in significant attenuation of the transmitted signal (named "0" state on the transmitted port).

We use simple coupled mode theory (CMT) to describe the dynamic tunneling and coupling between the photon densities in whispering gallery mode (WGM) and chaotic modes [21,31] (Supplementary file section S2):

$$\begin{pmatrix} \dot{a}_m \\ \dot{a}_{ch} \end{pmatrix} = \begin{bmatrix} -\left(\frac{1}{\tau_c} + \frac{1}{\tau_m}\right) + \kappa & -\kappa \\ -\kappa & -\left(\frac{1}{\tau_{ch}} + \frac{1}{\tau_m}\right) + \kappa \end{bmatrix} \begin{pmatrix} a_m \\ a_{ch} \end{pmatrix} + \begin{pmatrix} 0 \\ \frac{a_{in}}{\tau_c} \end{pmatrix} \quad (1)$$

Where $a_m$ and $a_{ch}$ represent the photon numbers in the WGM and chaotic fields, while $1/\tau_m$ and $1/\tau_{ch}$ are their optical decay rates, respectively. $a_{in}(t)$ is the complex input field varying rapidly with time (replicating the pulsed inputs). $1/\tau_c$ is the coupling rate between the input waveguide and the resonator. The parameter $\kappa$ defines the effective tunneling/transfer rate from chaotic modes to WGM. A value of $\kappa = 0$ implies no transfer rate between the states. The existence of silicon NCs, according to the results shown in Figure S4, potentially modifies $\kappa$ (Figure S4). With ultrafast rising edge, the energy conversion from chaotic fields to WGM (Figure S5a). Slower rising edge results in the majority of mode coupling to the chaotic mode (Figure S5b). The WGM has a low extinction ratio (output of '1' state), while the chaotic/high-order modes are close to the critical coupling point, and thus the transmitted photon intensity is significantly reduced ("0" state with slower rising edge).

The full field simulation replicates the pathway from chaotic mode to WGM in the NC MRR (Figure 2a-b). Figure 2a shows the representative snapshots of the resonator's spatial electric field intensity distribution from the three-dimensional Finite-Difference Time-Domain (3D FDTD) simulation. Firstly, a pulse signal couples from the bus waveguide into the higher-order modes in the microring resonator (the radius of the simulated ring is set to be 5μm as a toy model). Up to 39 picoseconds, the WGM occupancy increases up to 80% and stabilizes as WGM (Figure 2b). If we add random high refractive index structures (replicate the effect of NCs), the WGM portion firstly decreases, and the chaotic states dissipate as illustrated in Figure 2(a) (top panel of Figure 2a). Note that the existence of material nonuniformity (Raleigh scatterers of Si NCs) modifies the coupling between the chaotic fields and the WGMs (Supplementary Figure S4). Based on the coupled mode theory (CMT) simulations (equations 1-2 and Supplementary Section S2), the Si NC – SiN microresonator is likely to operate in the strong coupling regime, in which the coupling rate is higher than the difference between the chaotic mode and WGM dissipation rate, and results in oscillating fields between the two types of modes. The steady state solution of the CMP was used to fit such dynamics (Figure S6).

To eliminate the possibility of involving optical nonlinearities in the process, we next illustrate the time series of stochastic switching, in the format of output power versus input power, for a triangular input drive with intensity modulation from 8 μW to 200 μW. In the time-domain switching experiment, a triangular waveform with 50 measurements is shown in Figure 2(b). The output result reveals two distinct paths: the first path (blue) represents the signal coupling into and staying within the MRR, resulting in low output power ("0" state). The second path (red) corresponds to the signal coupling into the MRR but then dissipates, leading to high output power ("1" state). Note that while the on-state tracks in the input drive with high fidelity, the time series of the off-state shows temporal fluctuations in some of the off-cycles. We also emphasize the stability of our ring resonance and setup where the background thermal fluctuations are experimentally measured to be typically less than ~ 0.6 dB (and $\delta$ fluctuations of ~ 0.05), about an order of magnitude less than the observed on/off switching ratios (further detailed in Figure 5). We attribute this nondeterministic switching to the mode mixing and diffusion on the SiN – Si NC interface (Figure 2e-f).

This process is repeated on another independent experimental setup (Method). The second experiment setup involves a triangular waveform generator connected to a tunable laser from a different made, with varying clock frequencies ranging from 1 to 50 MHz. The output is collected by a lensed fiber and connected to a sampling oscilloscope (Figure 2g). The switching nature shows a dependence on the rising edge duration as we observed in Figure 1g (controlled through the clock frequency with the triangular waveform). At the rising edge duration of 0.5μs, the output shows a sharp transition from ~ 30% data mean value into a higher (~ 70%) data mean value. The clock frequency can thus be dialed as a method to bias-tune the on/off probability (Supplementary Section S3). The other evidence for the critical role of the rising edge is the waveform dependent switching possibility even at the same clock rate (Figure 1b). The fast-rising edge of the triangular waveform results in higher possibility of "1" state, while at the same clock rate, triangular and sinusoidal waveforms result in the "0" state.

We overlap the on/off stochastic switching ratio (*SSR*, extinction ratio between the two states measured on the oscilloscope) versus the input laser frequency detuning with the standard optical trans-mission spectra measurement by sweeping the wavelength of the input tunable laser. The hysteresis loops for two states (grey: '0 or tunneled states, red: '1' or dissipated states) are illustrated in the sub-panels. We note that the modulated stochastic switching is observed on *both* positive and negative detuned sides of the Lorentzian cavity. We note the extinction ratio/SSR for '1' and '0' states are maximized near the on-resonance, as the intensity of '0' state follows the transmission outline of the resonance. This is further detailed in Figure S2, along with the polarization dependence of the *SSR*. When driven with a periodic intensity-modulated waveform such as triangular, rectangular or sinusoidal waveforms, we observe the stochastic switching in the output transmission over a broadband range of parameters. To support the generality of the stochastic pattern generation, we performed sampling oscilloscope measurements of output waveforms *without* requiring any external electronic digital or pulse pattern generators (as shown in Figure 1b for rectangular, triangular, and sinusoidal waveforms). Here thousands of successive patterns are mapped, sampled at a 10 MHz input clock frequency. The return-to-zero eye diagram shows open eyes, illustrative of clear bistable switching for tunneling/dissipated states.

The random switching can also be replicated with different polarizations. The time domain series in Figure 3b shows a rectangular drive with 38 ns ramp up/downtime, 13 MHz repetition clock, and laser intensities from 5% to 100% peak power. Note that transverse electric (TE) and transverse magnetic (TM) polarized modes have different resonance wavelengths (Figure S1). The TE polarization (coupled into the thermally stabilized ring) shows the stochastic bistable on/off switch. At the same wavelength, the transverse magnetic polarization (not coupled to WGM) tracks the input drive precisely. Similarly, TM polarization (at a different input wavelength) exhibits a similar response. At the same wavelength, only the polarization coupled into the resonance exhibits a random switching response (Supplementary Section S4). A slightly different data mean value is observed between the two polarizations, likely due to the different overlap ratios of the many modes with nanocrystal regions. Similarly, after partially removing the nanocrystal, the dissipated state ('1' state) possibility significantly reduces, and the transient response replicates the typical MRR response (Figure 6).

The stochastic switching observed broadband from 1480 nm to 1560 nm. Figure 3c illustrates the cold cavity transmission of the ring (blue line). In this case, the critical coupling occurs around 1519 nm, as the intrinsic material absorption leads to propagation loss of ~ 6.8 dB/cm near the absorption peak and reduces to less than 2.3 dB/cm (4.3 dB/cm) at 1560 nm (1550 nm) [31]. Particularly we illustrate four example resonance modes (i through iv) on the return-to-zero eye diagrams. As shown in Figure 3d, we note the larger fluctuations in the bit switching at 1528.20 nm versus the longer wavelengths, resulting from the fluctuating energy level for '1' state and the noise from the additional amplifier. The extinction ratio of the diagram represents the on-resonance ring characteristics. In the inset of Figure 3c, we further illustrate the extinction ratio for the two states (eye diagrams in Figure 2d) correlates to the steady state measurement for a span of wavelength the telecommunication band laser covers.

The insensitivity of the switching statistics to the sampling threshold is an advantage of such a bistable system, which is implemented in a fully passive MRR structure. We applied a broad range of sampling thresholds for converting the analog waveform into binary digital bits (Figure 4a). The data mean value of output binaries maintains around 0.5+/-0.002 as the sampling threshold is selected from 35% to 68% to the peak power. The threshold variation range is 33% of peak power, compared to 0.16% tolerance for sampling threshold selection of a standard chaos system [11-12].

We quantify the statistics of the random bit generation, with the analyzed histograms of the random binary bit sequence shown in Figure S5. The time-bin of the stochastic switching is then simulated by random binary bits sequences, with probability density versus state duration for both on- and off-states. We obtain similar trends in the modeled $P_0$ and $t_0$ versus switch-on probability as in the measurements. We performed autocorrelation measurements of 16 Mbits time series data to examine its random nature further. The randomness can be directly observed by mapping into a bit pattern in a two-dimensional plane of 500×500 (Figure 4b). The 30 dB extinction ratio of the noise floor in Figure 4c also confirms the randomness, which can be further extended by increasing the number of data sets [35-36]. The random bit generation here is at 13 MHz, in the higher switch-on state probability (100 kHz clock frequency shows a different state bias, but the random nature is still present).

**Discussion**

We demonstrate chaotic mode in the multi-mode high $Q$ MRR, enabled by the low loss NCs clad SiN waveguide. The high refractive index contrast silicon NCs are densely embedded in the oxide gap in the MRR-waveguide coupling region, disturbing the delicate cavity build-up process. The physics-enabled bifurcation in the all-optical domain may replace sophisticated electronic processors. The passive devices can directly generate binary outputs with a clear 'eye', as captured by a standard sampling oscilloscope used in optical communications. We demonstrated for the first time chip-scale random number generation leveraging the random binary angular momentum tunneling through a chaotic state. The stochastic switching behavior is independent of modulated drive laser detuning, wavelength power, and polarization, but only the rising-edge dynamics of the input pulse and the overlap coefficient with the NC region. We studied large dataset distributions and validated them through the internal uncorrelated nature of secure communications and cryptography. Advanced functionalities, such as multichannel wavelength-division-multiplexed random bit generation, can be achieved on a single CMOS-processed ring resonator of ~ 400 µm$^2$ footprint. We envision that the compact and fully passive and integrated optical random pattern generator may enable a large-scale photonic reservoir computing system-on-chip, all-optical hardware implementation of diffusion models for regenerative AI [4,5,41].

Note that nonlinearity-induced chaos is extensively explored in micro-ring and other chip-scale micro resonators, induced by the interplay of different resonators or nonlinear responses [37-39]. Those chaotic responses and dynamics are highly sensitive to the input power and only appear on one side of the laser-cold cavity resonance detuning. In our device, symmetric stochastic switching appears symmetrically in both positive and negative tuning, which is delineated from the thermal and laser power fluctuation effects in high Q resonators, where thermal locking (whether for add- or drop-cavity resonances) occurs *only* one side of the detuning when the detuning fluctuates or is swept [40]. In our device, the output statistics and temporal evolution are independent of input power. Over 20dB dynamic range (input power from 5 µW to 500 µW), the switching dynamics are still observed with clear on/off states, elucidating the absence of optical nonlinearity effects on the stochastic nature (Figure 7).

We also measured the power spectral density (PSD) of the measured time series. The absence of frequency fluctuations validates that the stochastic nature does not arise from the pump laser, the coupling setup, or the supporting electronics. The sidebands of the amplitude noise are greatly lifted off the output from the device compared to the input clock signal, resembling the random nature of the output, while phase noise shows suppression. The non-deterministic and binary outputs are likely to be the chaotic ring amplified quantum phase noise from the laser source [7].

**Materials and Methods**

*Device fabrication and material characterization:* Plasma-enhanced chemical vapor deposition of 6500 Å SiN$_x$ is processed at a low temperature (350°C), with low interlayer stress. Deep-ultraviolet nanolithography defines the 1 µm width of the waveguides, followed by reactive etching and resist stripping. Then the silicon oxide is deposited on the waveguide with the hydrogen gas, and the silicon nanocrystals are formed in the silicon oxide matrix, seeded by the defects on the interface between PECVD silicon nitride and silicon oxide cladding layer. Photoluminescence spectra of the nanocrystal-ring samples are collected with the RENISHAW micro-Raman system under the 532 nm and 633 nm laser excitations, with 100× objective and 1 µm spot sizes (Inset of Figure 1c). The TEM sample was prepared by selectively removing the silicon substrate through wet etching. The chiplet was soaked in sodium chloride solution overnight, leaving the top device layer (silicon nitride and silicon nanocrystal) between the oxide layers. The thin film was then transferred to TEM grid for inspection. JEOL ARM200CF (operating at 200 kV, equipped with double aberration correctors and Gatan energy filter system) was used to obtain the high-resolution images of the silicon NCs.

*Optical measurements:* Stochastic bistable switching is measured and verified on two independent setups. Periodically modulated continuous-wave radiation generated from the tunable laser (Ando AQ4321) is sent onto the chip through a polarization controller and lensed fiber. A built-in power control circuit maintains the optical output intensity stability to within +/- 0.01 (0.05) dB or less over 5-minute (1-hour) durations. An integrated spot size converter reduces the total fiber-chip-fiber loss to 14 dB. Output transmission is collected by a power meter and a high-speed infrared photo-receiver (New Focus 1554B, DC-12GHz bandwidth). The fast photodetector is connected to the digital phosphor oscilloscope (Tektronix TDS 7404, DC-4GHz bandwidth). For the return-to-zero eye diagram measurement in the second setup, the tunable laser (Santec TSL), driven by a function generator (HP8116A), sends out different waveforms into the device through a polarization controller (HP11896A). The output signal is amplified by EDFA (Photonetics Fiberam-BT19) with the tunable filter before sending it into a wide-bandwidth sampling oscilloscope (Agilent 86100A with module 86109B).

*Optical simulations:* The full field simulations are performed by the 3D FDTD method [42]. The conformal mesh with a spatial resolution less than 1/10 of the local feature size is applied for FDTD simulation. The radius of the microring was set at 5 µm. The center wavelength of the input light source was set at the resonance wavelength, and the duration varies between a few fs to 750ps for resolving the pulse dynamics dependent chaotic mode response. The time-varying mode coupling portion is obtained through fitting the FDTD simulation with the CMT model.

**Acknowledgments**

The authors acknowledge the fruitful discussions with Senghui Kim, Profs. Xuefeng Jiang and Giulia Galli. The PL mapping is performed in Dr. T. Heinz's lab with assistance from Dr. Y. Li. The high-temperature annealing (for eliminating non-Si-NCs effects such as H-N bonds) postprocessing is performed in Dr. J. Hone's lab with assistance from Xian Zhang. The TEM work was supported by the U.S. Department of Energy, Office of Basic Energy Science, Division of Materials Science and Engineering, under Contract No. DE-SC0012704. This work is sponsored by the National Science Foundation with grant no. 2338546.

**Figures**

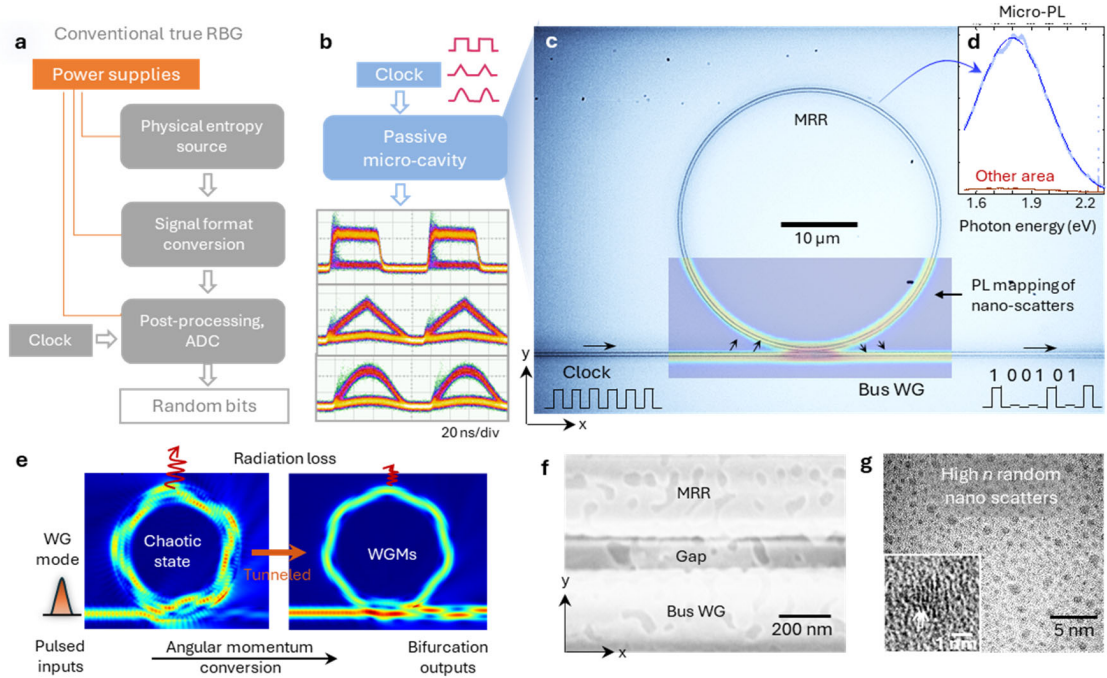

**Figure 1. Silicon nanocrystal perturbed momentum transfer in a passive microring resonator (MRR) towards a chip-scale binary random output. a,** System schematic diagram of conventional true random bits generator (RBG). The analog random signal is generated from the physical entropy source, converted or amplified before reaching the postprocessing unit for analog-to-digital conversion (ADC). **b,** Physical implementation of fully passive microsystem directly converts input clock signal to clock-wave form defined output binary pattern. Bottom inset: Measured return-to-zero eye diagram of chip-scale random bit pattern directly generated from the device under test. Measurements for rectangular, triangular, and sinusoidal waveform drive at zero detuning, each with the random open-eye pattern generation. **c,** Optical microscope image of PECVD-grown silicon nitride ring resonator with 40 µm diameters. The densely packed Si nanocrystals are embedded in $SiO_2$ cladding to the SiN interface. Middle inset: micro-photoluminescence intensity spatial map (photon energy of 1.8eV) superimposed on the device micrograph, with **d,** typical micro-PL spectra with top excited laser spot (~300nm) focused on top of the waveguide. **e.** Short-time snapshots for the full field numerical simulation of the mode, evolving from the waveguide excitation, propagation of chaotic states (left), tunneling to the WGM (right), or dissipation to radiative continuum. **f.** SEM of the partially etched MRR - waveguide coupling region, exposing non-uniform and deep subwavelength voids hosting silicon nanocrystals in the oxide cladding close to the SiN layer. **g.** High-resolution transmission electron micrograph (TEM) of silicon nanocrystal (Si NCs) ensemble embedded in the oxide cladding around the silicon nitride waveguide.

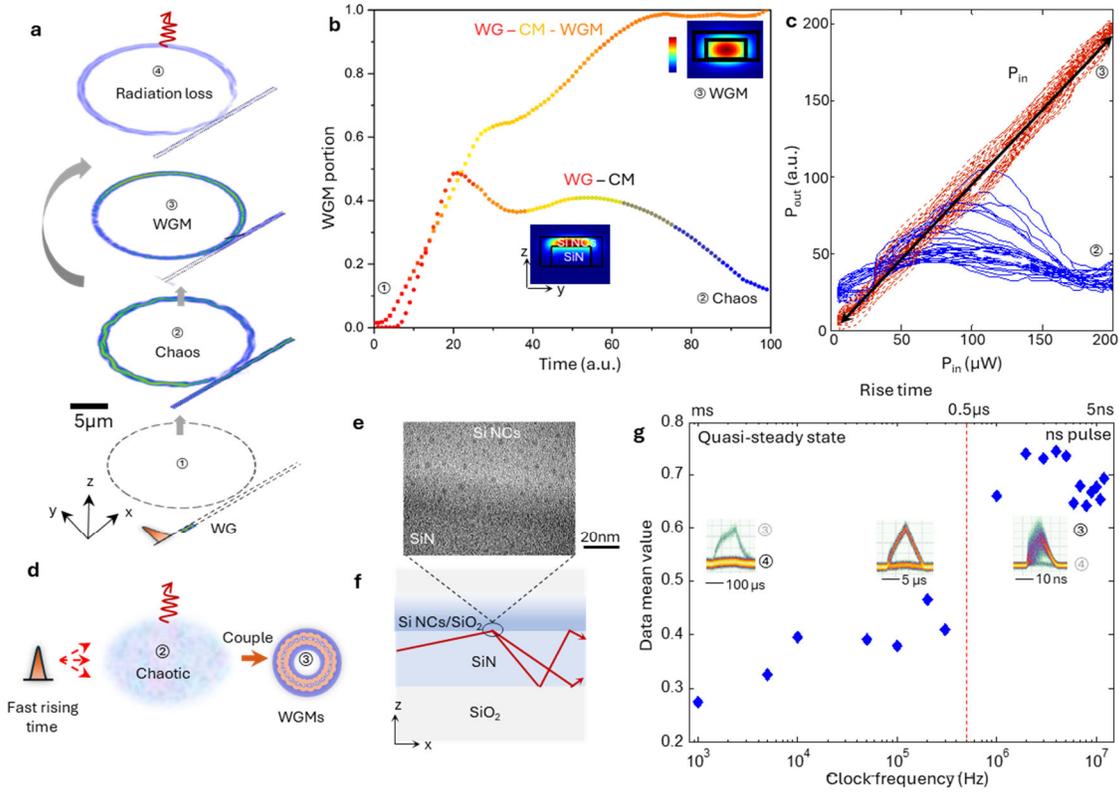

**Figure 2. Transient dynamics of the stochastic binary tunneling**. **a,** Short-time snapshots for the full field numerical simulation of the mode, evolving from the waveguide excitation, coupling to chaotic channel, propagation of chaotic states, tunneling to the WGM, or dissipation to radiative continuum. **b,** Temporal evolutions of excited WGM portion derived from spatial intensity distribution and two-level modeled value for both tunneled and dissipated paths through chaotic states. Insets: Exemplary cross-sectional mode profiles in the Si NC clad SiN WG for chaotic mode and WGM mode. **c,** measured time-domain switching for "1" (red) and "0" (blue) states, through a real-time oscilloscope with triangular waveform input. 50-period time series are plotted to illustrate the metastable fluctuations. **d.** System schematics of non-deterministic tunneling in the micro-resonator supporting chaotic modes. **e,** TEM of the interface between silicon oxide and SiN interface. **f,** Schematic diagram of the mode diffusion and mixing on the SiN-NC interface. **g,** Measured tunneling or radiation possibilities (data mean value) versus repetition rate (clock frequency) of the input triangular waveform drive, with clock frequencies from 1 to 50 MHz, for $P_{in\_max}$ of 100 μW. Inset: sampling oscilloscope measured eye diagram. A faster-rising edge results in higher chaotic tunneling possibilities.

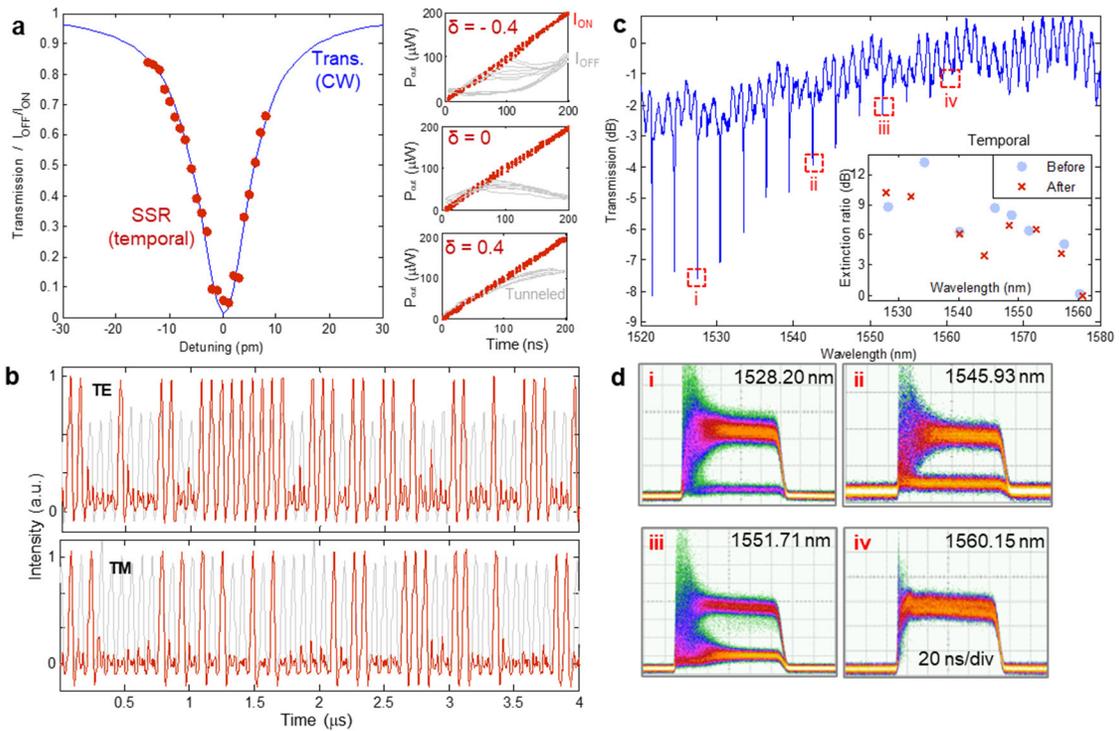

**Figure 3. Binary random pattern generation. a**, measured temporal stochastic on/off switching ratios for *both* blue and red-side detuning from the cavity resonance (red dots). A resulting cavity filter lineshape is experimentally superimposed on the broadband stochastic switching. The time domain measurements of the on/off intensity ratio (red dots) match the measured spectral transmission lineshape (solid blue line). Right panels are traces of the stochastic switching dynamics, for example, laser-cavity detuning on both sides [$\delta = (\lambda_{laser} - \lambda_{cavity})/\Delta\lambda$]. b. Output time-series transmission of Si-NC clad SiN ring resonator under periodic waveform drive. The upper panel compares the transverse electric (TE; red) versus the transverse magnetic (TM; grey) polarized light on resonance with the Si-NC clad SiN. The lower panel shows the TE polarized input drive at 12 pm (red) and zero (grey) detuning. **c,** Transmission of 20 μm radius ring resonator. Inset: extinction ratio ($I_{ON}/I_{OFF}$) of the return-to-zero eye diagram of output from 1528 nm to 1560 nm, before (empty circles) and after (solid circles) 1150°C annealing for 1 hour in Ar environment. **d,** Measured eye diagrams for resonances at 1528.20, 1545.93, 1551.71, and 1560.15 nm. Input pulse repetition rate is 10 MHz.

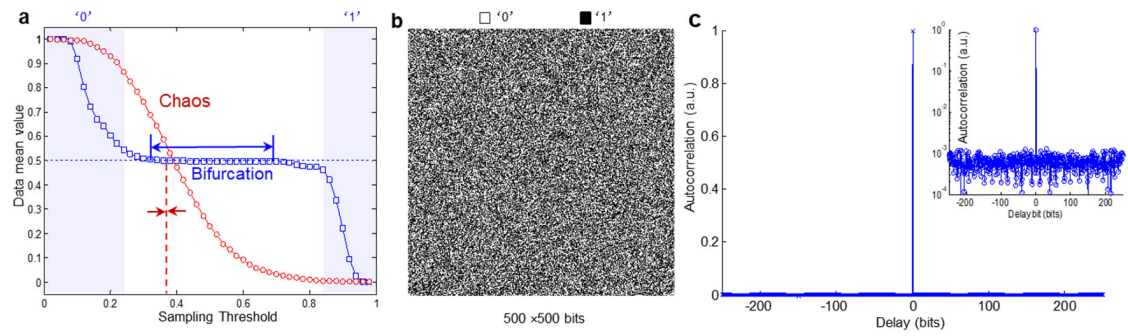

**Figure 4. Bifurcation outputs and verification of random outputs. a,** Data mean value versus sampling threshold for the measured data (blue), compared to a typical chaotic source (red). **b,** Random bit patterns in a two-dimensional plane. Bits 1 and 0 are converted into black and white dots and placed from left to right (from top to bottom); 500×500 bits are shown. **c.** Autocorrelation functions of the output random bits on a linear scale. Inset: log scale.

**Extended data**

The resonance fluctuation is majorly induced by room temperature fluctuations and relative refractive index change. The resonance detuning fluctuation leads to on-resonance transmission shifts between 0.8dB, which is an order of magnitude lower than the random switching we observed, at much slower rate (~min).

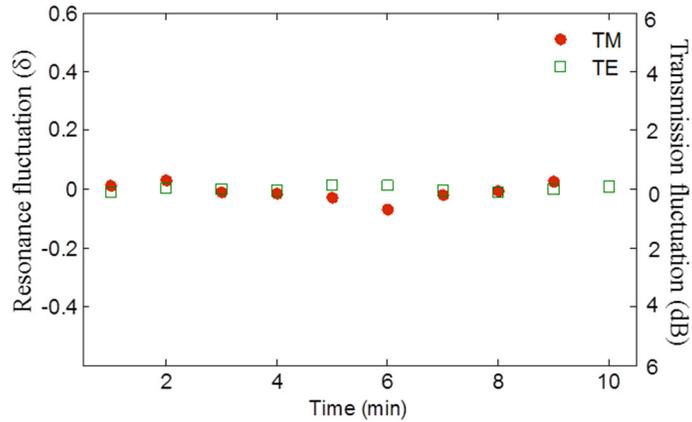

**Figure 5**. **Measured stability of ring resonance and transmission, for an example 1532 nm resonance, over 10-minute period, for TM (red dots) and TE (green empty squares) respectively.** At room temperature, the transmission for the laser on resonance with the ring fluctuates less than 0.8 dB with a 0.3 dB standard deviation, which is almost negligible for ±6 dB observations of stochastic switching between on- and off-states. The normalized detuning $\delta$ [defined as $(\lambda_{laser} - \lambda_{cavity})/\Delta\lambda$] fluctuates on an average of 0.03.

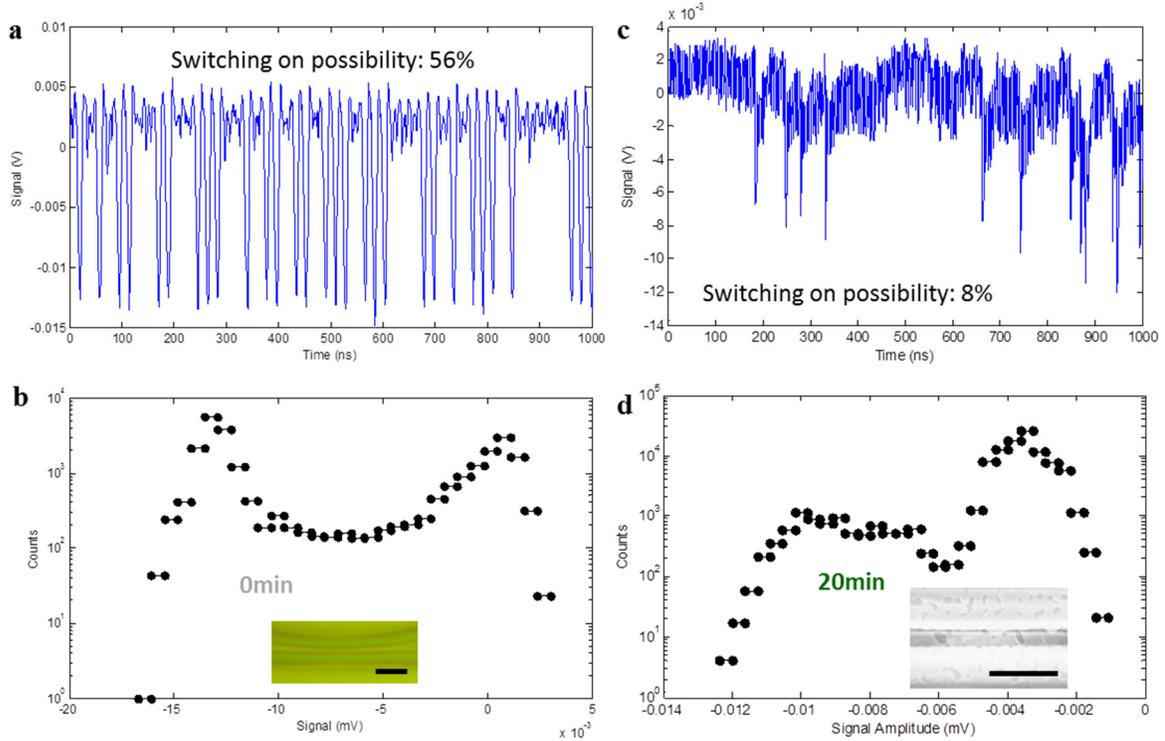

**Figure 6. Reduce the switching possibility of the device with mostly removed silicon NCs.** (a) The time domain switching dynamics under 13MHz periodic excitation; (b) Histogram of $10^6$ bits in (a). Inset: top optical view of the coupling part between the waveguide and the ring. Scale bar: 2 μm. (c) Switching dynamics of the device after removing most of the top oxide. A thin layer oxide (<50nm) was left to avoid over-etching. The input periodic pulse is the same as (a). (d) Histogram of (c). Inset: top SEM image. Scale bar: 2 μm. The dissipation (non-tunneling) possibility reduces from 56% to 8% and 0% for non-etched, partially removed and totally removed NCs samples, respectively.

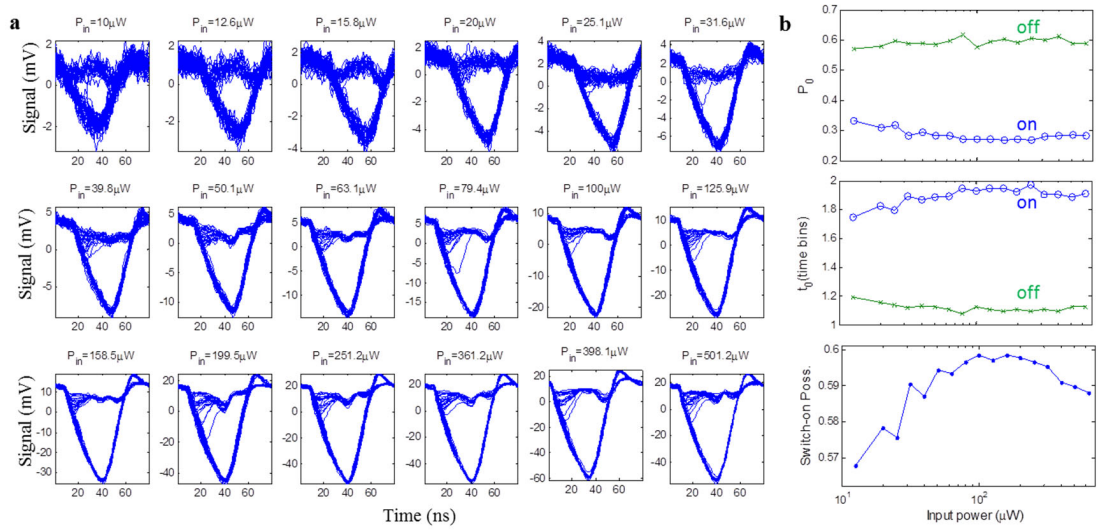

**Figure 7. Power independence of the stochastic switching.** (a) Superposition of 72 single shots, as the peak of the input power increases from 10μW to 0.5mW. (b) Parameters of the statistical distribution (amplitude and lifetime) are shown in (b), including *P0*, *t0* and switch on possibility. After synchronizing with input clock, the output distribution follows exponential decay: $P=P_0 e^{-t/t0}$

Supplementary Information for

Bistable random momentum transfer in a linear on-chip resonator

Tingyi Gu[1,2†*], Lorry Chang[1†], Jiagui Wu[2,3], Lijun Wu[4], Hwaseob Lee[1], Young-Kai Chen[5], Masudur Rahim[1], Po Dong[5], and Chee Wei Wong[2,3]*

[1] Department of Electrical Engineering, University of Delaware, Newark, DE 19711, United States

[2] Columbia University, New York, NY 10027, United States

[3] Fang Lu Mesoscopic Optics and Quantum Electronics Laboratory, University of California, Los Angeles, CA 90095, United States

[4] Condensed Matter Physics and Materials Science Department, Brookhaven National Laboratory, Upton, New York 11973, United States

[5] II-VI Incorporated, 48800 Milmont Drive, Milmont, California 94538, United State

[†]Authors contributed equally, * correspondent authors

## S1. Optical and nanostructure characterization of the device

$SiN_x$ films are grown by plasma-enhanced chemical vapor deposition (PECVD). The $SiH_4$ and $N_2$ gas flow rates are 80 sccm and 4000 sccm, respectively. 400W RF power was applied to reach the SiN deposition rate of 22.8A/s. The growth temperature is kept at 350⁰. The refractive index of SiN film is measured to be 2.03 [SR1]. After the UV lithography and etching for the formation of waveguide structures, a PECVD $SiO_2$ layer was deposited on top with gas flows of $SiH_4$ and $O_2$.

Since no annealing was performed during the thin film growth, we believe those silicon nanocrystals (NCs) embedded in the oxide matrix were formed through gas-phase synthesis with plasma enhancement, with Hydrogen diffused from PECVD SiN waveguide at the growth temperature of 350⁰. Since hydrogen is critical for the formation of Si NCs in the oxide matrix, the area away from SiN surface does involve sufficient H in the oxide growth process and thus Si NCs were absent (more details in Fig. S2).

The linear loss of the waveguides is measured by the sampling of different channel waveguide length samples, from 25 mm to 44 mm with a 5 mm step. The fitted transmitted power versus waveguide length with the linear model gives a maximum propagation loss of ~ 6.8 dB/cm at 1520 nm, which is reduced to 4.3 dB/cm at 1550 nm, and further reduced to less than 2.3 dB/cm at 1560 nm. The fiber-to-fiber coupling loss from the two facets is ~ -13.5 dB.

We measured three micro-rings with diameters of ~20, 40 and 70 μm, with loaded, intrinsic quality factor ($Q$) and FSR respectively of 24,500, 49,000, 8.7nm; 69,600, 175,000, 4.4nm; and 77,300, 244,000, 2.9nm. The intrinsic $Q$ is inversely related to the total loss rate of the resonator, including material absorption loss and scattering loss. The absorption from the chemical bonds (second harmonic absorption of H-N bond near 1520nm) is wavelength dependent while the scattering of the silicon nanocrystals is broadband. Smaller rings exhibit lower $Q$ due to additional bending loss. With the assistance of coupled mode theory analysis, we extract the broadband scattering loss of 1dB/cm, and the maximum linear loss originating from the H-N bond absorption (near 1520 nm) adds another -2.8dB/cm [SR2].

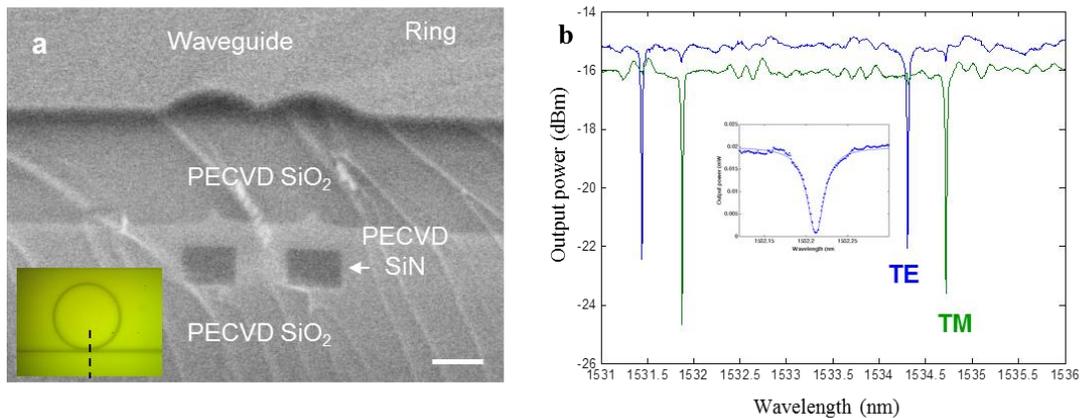

**Figure S1 | Structure and linear optical properties of the device. a,** SEM of the cross-section of the ring resonator. 650nm PECVD silicon nitride is sandwiched between the PECVD silicon oxide up cladding layer and the thermal oxide lower cladding layer. Scale bar: 1um. Inset: Optical image of top view of the ring, where the dashed line shows the cleaved position for the SEM image. **b,** Output spectrum of TE and TM polarized input with 0dBm input power. Inset: Transmission spectrum of measured modes with a 16 pm linewidth and 14 dB extinction ratio.

All of microresonators across the wafer show random switching behavior, with an extinction ratio and average switching possibility varying with wavelength and radius. The cross-section of the ring-waveguide coupling is shown in Fig. S1a, where the PECVD silicon oxide surrounds PECVD silicon nitride. The high aspect ratio of the silicon nitride waveguide (1μm wide, 0.65μm tall) distinguished TE and TM modes (as in Fig. S1b). The quality factor can be read by using coupled mode theory to fit the transmission spectrum of one of those modes. The largest ring with 70 μm diameter with a ring-waveguide gap of 0.5μm gives FWHM of ~16pm.

Note that the photoluminescence signal can be attributed to either the defects in SiN or silicon nanocrystals [SR33-SR7]. To track the origin of PL (from the core SiN waveguide or cladding layer), we performed the multi-step wet etching and inspection cycles. The PL intensity reduces with the removal of the top oxide layer (hosting the Si nanocrystals) (Fig. S2).

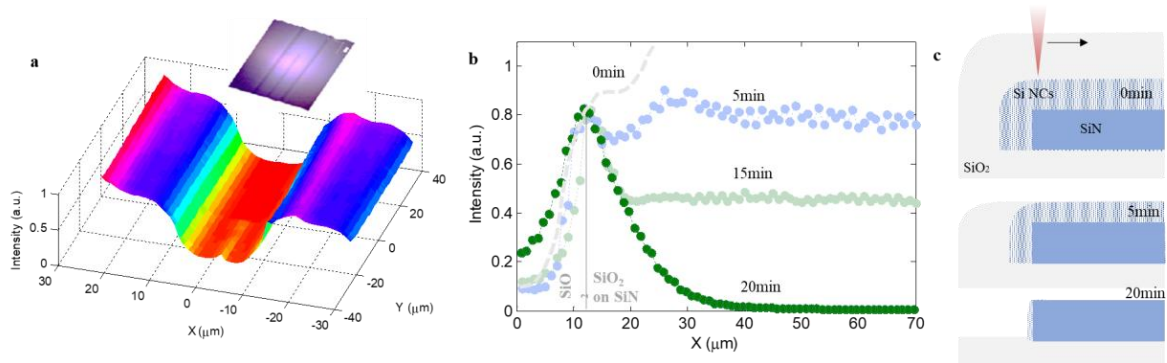

**Figure S2 | Confirmation of silicon nanocrystals distribution in cladding layer by wet etching.** (a) The photoluminescence intensity mapping on the tip of the waveguide (Inset: optical image). The part with silicon nitride (two sides and middle waveguide) shows a much stronger PL signal than the part without SiN. (b) Map of PL intensity along X axis after 50% BOE wet etching of 0 min (grey dashed line), 5 min (blue dotted line), 15 min (light green dotted line), and 20 min (dark green dotted line). The edge of the SiN is marked in the figure at X=12μm. (c) Schematic device cross-section with different oxide wet-etching time (corresponds to b). The scanning micro-photoluminescent mapping (results in a and b) was obtained through the top incident/collected focused beam.

The additional cladding layer of Si NCs significantly alters the number of supported modes and their effective index in the waveguides (Fig. S3).

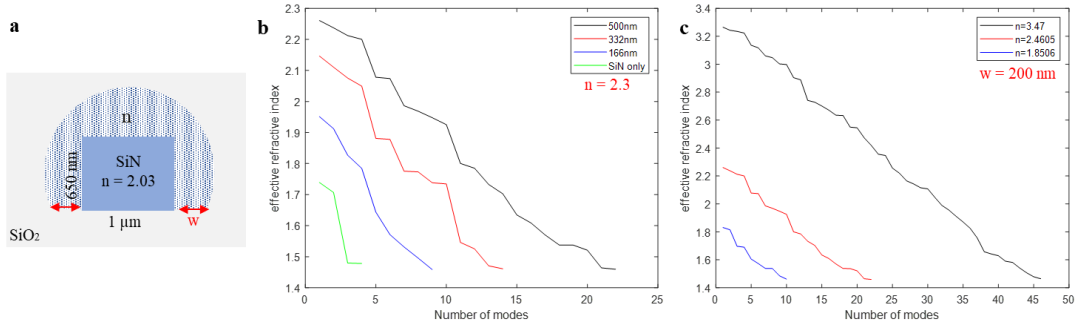

**Figure S3 | Waveguide cross-section and effective index of the supported modes.** (a) Cross section of the SiN waveguides. The refractive index of the PECVD-grown SiN is measured to be 2.03 [SR1]. (b) Effective index versus supported modes for increasing thickness of the Si NCs embedded oxide layer. (c) With a fixed Si NC embedded oxide layer, increasing the effective index (scales with Si NCs density) results in a higher effective index in many more modes.

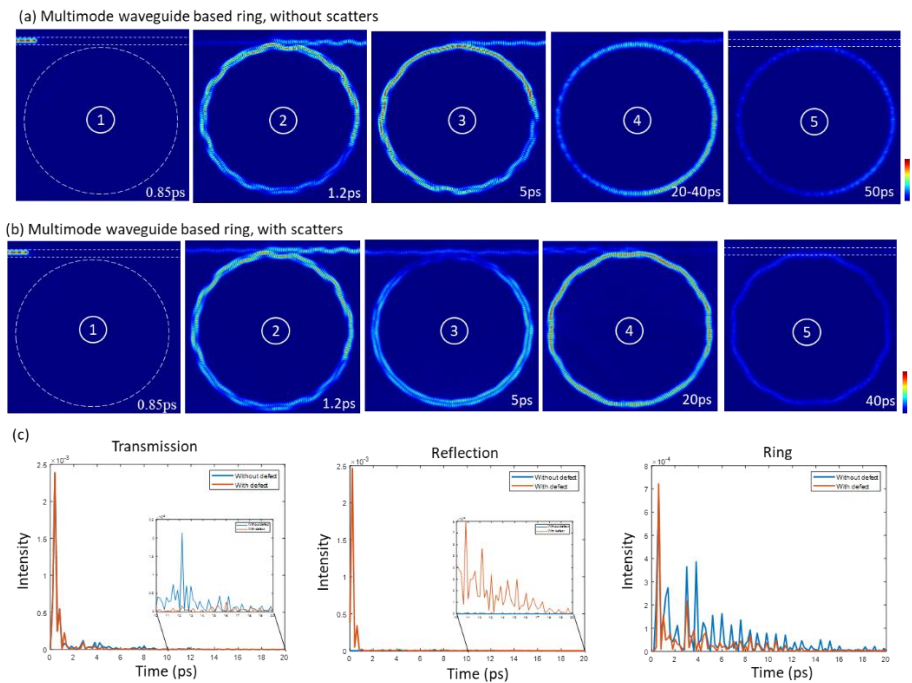

**Fig. S4 | Impacts of scatters/defects on the chaotic – WGM mode transitions.** (a) Mode dynamics in the multi-mode microresonator without and (b) with deep subwavelength scatters/defects. (c) The intensity dynamics for the monitors placed on the input port, transmitted port and within the resonator, comparing the results without (red) and with scatters (blue). Ring radius: 5 µm.

Then we conducted the 3D FDTD numerical simulations in the multi-mode micro-ring with and without scatters. The first set focuses on the impact of the scatters. A 750 ps pulse excitation was injected into the structure through the bus waveguide (labeled 1 in Fig. R2). As time progresses, chaotic states emerge within the micro-ring, apparent from the unevenly distributed modes around the ring (labeled 2). After a few picoseconds, the occupancy of WGMs increases, and the modes stabilize towards WGMs (labeled 3-4). Finally, the mode dissipated through the radiation/intrinsic loss (labeled 5). Significant difference is observed in stage 3 and 4, after adding the scatters. The forward and backward scattering changes the direction of input light and modifies the mode conversion dynamics in the resonator.

**S2. Dynamic tunneling through chaotic states**

Parallel to the 3D FDTD simulation, we build analytical modes of coupled mode theory (CMT) to describe the dynamic tunneling and coupling between the photon densities in whispering gallery mode (WGM) and chaotic modes [SR8, SR9]:

$$\frac{da_m}{dt} = -\left(i(\omega_L - \omega_0) + \frac{1}{\tau_c} + \frac{1}{\tau_m}\right) a_m - \kappa(a_{ch} - a_m) \tag{S1}$$

$$\frac{da_{ch}}{dt} = -\left(i(\omega_L - \omega_0) + \frac{1}{\tau_c} + \frac{1}{\tau_{ch}}\right) a_{ch} - \kappa(a_m - a_{ch}) + \frac{a_{in}(t)}{\tau_c} \tag{S2}$$

Where $a_m$ and $a_{ch}$ represent the photon numbers in the WGM and chaotic fields, while $1/\tau_m$ and $1/\tau_{ch}$ are their corresponding optical decay rates. $\omega_L$ is the angular frequency of the input laser, and the $\omega_0$ is the one for the cold cavity resonance. $a_{in}(t)$ is the complex input field varying rapidly with time (replicating the pulsed inputs). $1/\tau_c$ is the coupling rate between the input waveguide and the resonator. The parameter $\kappa$ defines the effective tunneling/transfer rate from chaotic modes to WGM. A value of $\kappa = 0$ implies no transfer rate between the states. The existence of silicon NCs, according to the results shown in Fig. S4, potentially modifies $\kappa$.

Here we plot the mode dynamics for inputs with different rising edges. Here we set the system to be weak-coupling regime ($2\kappa < \frac{1}{\tau_{ch}} - \frac{1}{\tau_m}$). With ultrafast rising edge, the energy conversion from chaotic fields to WGM (Fig. S4a). Slower rising edge results in the majority of mode coupling to chaotic mode, as shown in the rising chaotic mode intensity in Fig. S4b). Similar dynamics are shown within the strong coupling regime ($2\kappa > \frac{1}{\tau_{ch}} - \frac{1}{\tau_m}$). In strong coupling regime, the fields oscillate between the WGM and Chaotic fields, with energy exchange rate increase with $\kappa$. The oscillation between WGM and chaotic modes within the strong chaotic mode-WGM coupling region, which is likely to be the origin and nondeterministic final states.

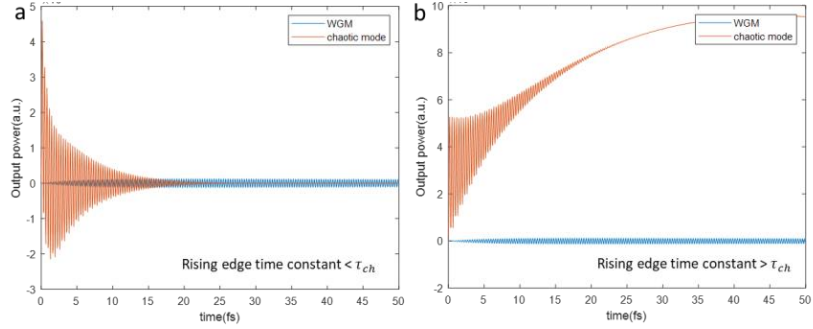

**Fig. S5 | Cavity field dynamics in chaotic mode (orange) and WGM (blue) under excitation with different rising edge constants**. (a) Rising edge faster (b) slower than the photon lifetime of chaotic mode.

Once the system is stabilized after the fast-rising edge, the WGM proportion ($\eta(t)$) can be expressed as.

$$\eta(t) = \frac{n_m}{n_o + n_m} = \frac{2\kappa}{\gamma_m - \gamma_o + 2\kappa + \sqrt{(\gamma_o - \gamma_m)^2 + 4\kappa} cothcoth\left(\frac{1}{2}t\sqrt{(\gamma_o - \gamma_m)^2 + 4\kappa}\right)} \qquad (S3)$$

Where $\eta(t)$ is the WGM proportion with respect to time. Using equation (S-3), we fit our simulation data and extracted the parameters: $\gamma_m = 0.009$, $\gamma_c = 0.02$, and $\kappa = 0.0005$. The extracted parameters align well with our simulation results. The chaotic field decay rate is faster than the WGM decay rate, which is expected since chaotic fields dissipate more quickly over time, leaving only the WGM port after their dissipation. Furthermore, the small value of suggests minimally effective tunneling between the two states.

$\eta(t)$ can also be obtained from the full field numerical simulations. By placing a waveguide cross-sectional monitor in the ring, we retrieved the instantaneous WGM portion by using Equation 1 [S8]. The WGM portion is calculated by projecting the instantaneous electric field onto the established WGM electric field. This allows us to quantify the portion of the mode converted to the WGM during the process:

$$\eta_{WGM}(t) = \left| \frac{\int E(t;x,y,z) E^*_{WGM}(x,y,z) dxdydz}{\sqrt{\int |E(t;x,y,z)|^2 dxdydz \int |E_{WGM}(x,y,z)|^2 dxdydz}} \right| \quad (S4)$$

Where $\eta(t)$ is the WGM proportion of the mode, which evolves rapidly with time. $E$ is the transient complex electric field in the specific location ($x,y,z$), while $E_{WGM}$ is the corresponding field of WGM mode at the same location.

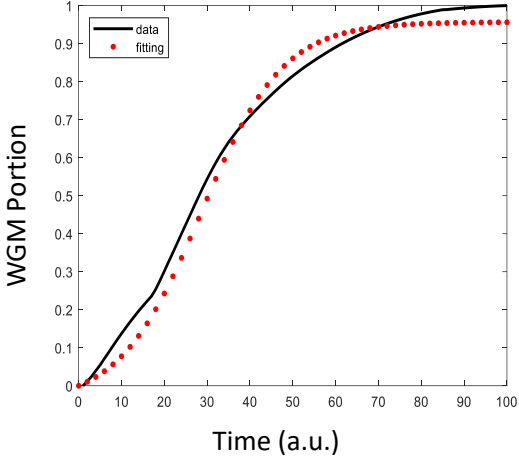

In Fig. S6, we compare the numerically extracted $\eta(t)$ (from equation S4) and the fitting with analytical model (equation S3).

**Figure S6 | Coupled mode theory fitting for the chaotic mode – WDM conversion.** Dark solid curve: numerical simulation. Dots: two-state CMT models in equation S1.

## S3. Time-bin characterization of stochastic switching dynamics

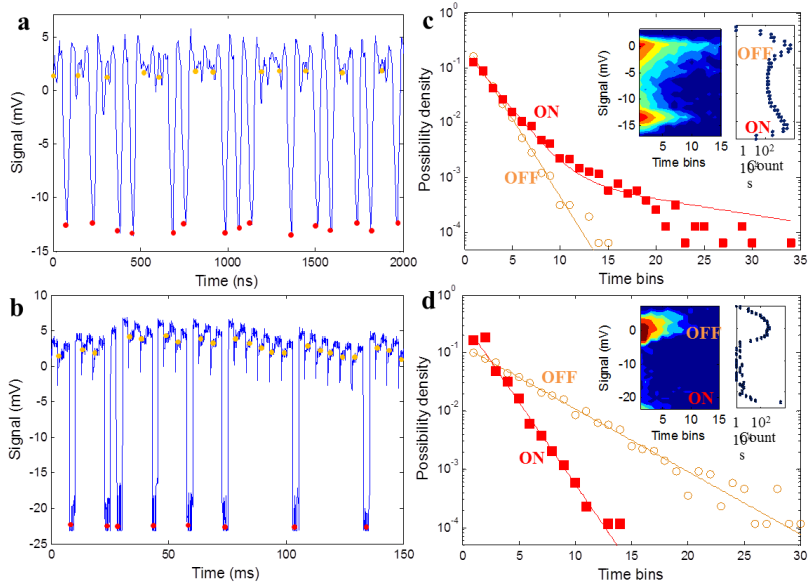

**Figure S7 | Stochastic switching dynamics with periodic perturbation, for the ring with 70um radius.** The temporal response of the bistable system to the modulated laser signal with peak power at 0.4mW and chopping frequency at (a) 13MHz (b) 200Hz. The red and yellow dots mark points sampled for the 'off' and 'on' states. The laser is set on resonance with the ring resonance at

S-7

1484.57nm. (c) and (d) are the ON and OFF time possibility distribution for (a) and (b) respectively. The total data points are 16,000 for both of the plots. The insets are the output signal collected on photodetector versus time bins (left) and histograms for both states. The distribution follows exponential decay: $P=P_0 e^{-t/t_0}$ (c) f=13MHz: $P_{0\_ON}$=0.130, $T_{0\_ON}$=2.01, $P_{0\_OFF}$=0.165, $T_{0\_OFF}$= 1.53; (d) f=200Hz: $P_{0\_ON}$=0.196, $T_{0\_ON}$=1.53, $P_{0\_OFF}$=0.103, $T_{0\_OFF}$= 4.03.

The switch-on possibility jumps from 0.2 to 0.8 when the detuning is less than 1. The laser repetition period is adjusted between 76 ns to 5 ms. Stochastic switching is observed under both conditions as shown in Fig. S5a-b, but the statistics, including possibility density versus time bins for both ON and OFF states, are significantly different (Fig. S5c-d).

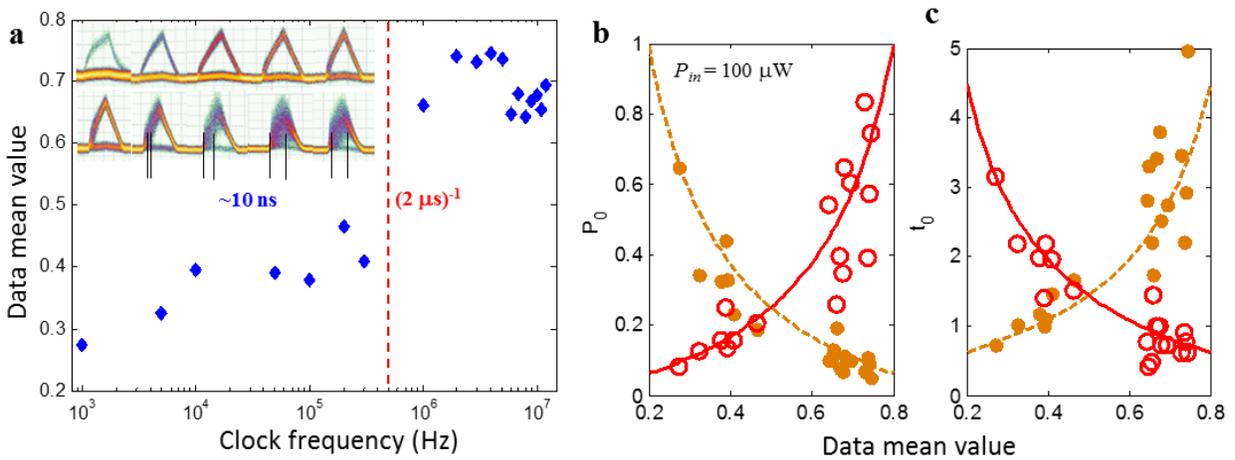

**Figure S8 | Laser modulation frequency dependent data statistics. a,** Switch-on possibilities (data mean value) versus clock frequency for the device with rectangular waveform drive, with clock frequencies from 1 to 50 MHz, for $P_{in}$ of 100 μW. Inset: eye diagram for triangular waveform drive. **b** and **c,** Amplitude $P_0$ and lifetime $t_0$ versus switch-possibility for the on- (red) and off- (orange) states. The red crosses are experimental data for on-states, and the orange dots are for off-states. Statistical models for on- and off-states are illustrated in the red solid and the orange dashed curves respectively.

### S4. Detuning and polarization dependence of the statistical binary outputs

For examining the resonance behavior of the random switching, we measure both frequency domain dependence of extinction ratio and time domain SSR. We see the clear detuning and polarization-dependent SSR resembling the ring transmission measured at CW laser input. It excludes the involvement of optical nonlinearity from the light-ring interaction in the mechanism of random switching (Fig. S7). The polarization and detuning-dependent PSD are plotted accordingly.

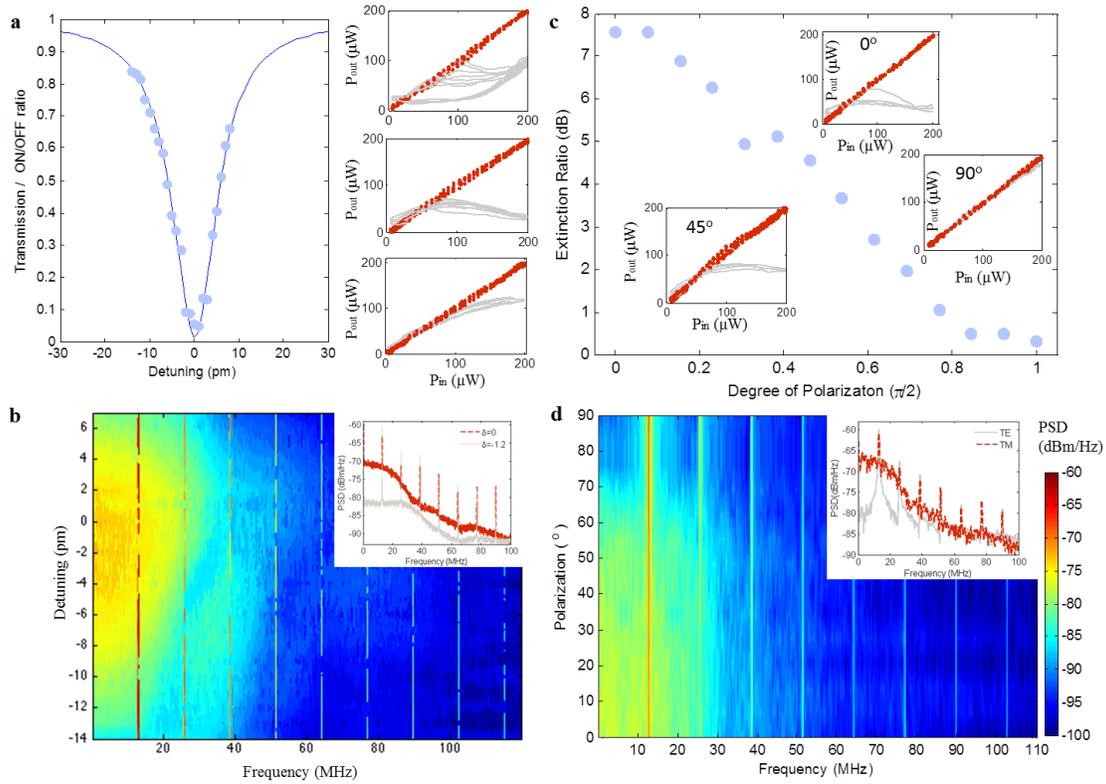

**Figure S9 | Detuning and polarization-dependent signal levels of on and off states** (a) Transmission spectrum (Solid blue line) of a single resonance located at 1493.104nm with 10.66pm FWHM (middle). The hysteresis loop for the different laser-cavity detunings at -4, 0, and 4 pm (subplot 1-3). The light blue dots are experimental measurements, and the blue curve is the coupled mode theory curve fitting). (b) Correspondent power density for different detunings as shown in a. Inset: the PSD for 0 detuning (red) and -1.2 detuning (grey) (c) Extinction ratio between ON and OFF states as the on-resonance laser polarization tuned from TE to TM in step of 7°. Insets: the hysteresis loop for TE (0°), TM (90°), and in between (45°). (d) Correspondent power density spectrum for TE to TM polarized light, as shown in (c). Inset: PSD for TE and TM polarized light.